%% file: Shi_metalpoor.tex
\documentclass{natureprintstyle}

\usepackage{epsfig}
\usepackage{color}
\usepackage{bm}
\usepackage{graphicx}
\usepackage{longtable}
\usepackage{amssymb}
\usepackage{rotating}

\newcommand{\apj}{Astrophys. J.}

\newcommand{\apjs}{Astrophys. J. Supp.}
\newcommand{\araa}{Annu. Rev. Astron. Astrophys.}
\newcommand{\mnras}{Mon. Not. R. Astron. Soc.}

\newcommand{\aap}{Astron. Astrophys.}
\newcommand{\aj}{Astron. J.}
\newcommand{\nat}{Nature}

\newcommand{\aaps}{A\&AS}

\title{Inefficient Star Formation In Extremely Metal Poor Galaxies}

\author{Yong Shi$^{1,2}$, Lee Armus$^{3}$, George Helou$^{3}$, Sabrina Stierwalt$^{4}$, Yu Gao$^{5}$, Junzhi Wang$^{6}$, Zhi-Yu Zhang$^{7}$, Qiusheng Gu$^{1,2}$}
\begin{document}

\maketitle

\let\thefootnote\relax\footnote{
\begin{affiliations}
  \item School of Astronomy and Space Science, Nanjing University, Nanjing 210093, China.
  \item Key Laboratory of Modern Astronomy and Astrophysics (Nanjing University), Ministry of Education, Nanjing 210093, China.
  \item Infrared  Processing and Analysis   Center,   California    Institute   of   Technology,   1200 E. California Blvd, Pasadena, CA 91125, USA.
  \item Department of Astronomy, University of Virginia, P.O. Box 400325, Charlottesville, VA 22904, USA.
  \item Purple Mountain Observatory,  Chinese Academy of Sciences, 2 West Beijing Road, Nanjing 210008, China.
  \item Shanghai Astronomical Observatory, Chinese Academy of Sciences, 80 Nandan Road, Shanghai 200030, China.
  \item Institute for Astronomy, University of Edinburgh, Royal Observatory, Blackford Hill, Edinburgh EH9 3HJ, UK
\end{affiliations}
}
\vspace{-3.5mm}
\begin{abstract}

The first  galaxies contain stars  born out of  gas with little  or no
metals.   The lack  of metals  is  expected to  inhibit efficient  gas
cooling  and  star  formation\cite{Ostriker10,  Krumholz13}  but  this
effect  has yet  to  be  observed in  galaxies  with oxygen  abundance
relative to hydrogen below a tenth of that of the Sun\cite{Krumholz13,
Bigiel08, Bolatto11}.  Extremely metal poor nearby galaxies may be our
best  local laboratories for  studying in  detail the  conditions that
prevailed in low metallicity galaxies at early epochs. Carbon Monoxide
(CO)   emission   is   unreliable   as   tracers   of   gas   at   low
metallicities\cite{Elmegreen13,  Bolatto13, Leroy11},  and  while dust
has     been    used     to    trace     gas     in    low-metallicity
galaxies\cite{Elmegreen13,     Fisher14,     Hunt14,    Remy-Ruyer14},
low-spatial resolution in the  far-infrared has typically led to large
uncertainties\cite{Hunt14,    Remy-Ruyer14}.      Here    we    report
spatially-resolved infrared  observations of two  galaxies with oxygen
abundances  below 10 per  cent solar,  and show  that stars  form very
inefficiently in seven star-forming clumps of these galaxies. The star
formation efficiencies  are more  than ten times  lower than  found in
normal,  metal rich  galaxies today, suggesting that star
formation may have been very inefficient in the early Universe.

\end{abstract}

The two  galaxies that are  the focus of  this study are Sextans  A, a
dwarf   irregular  at   1.4   Mpc  with   oxygen   abundance  of   7\%
Solar\cite{Pettini04,     Kniazev05},    and     ESO     146-G14,    a
low-surface-brightness  galaxy  at  22.5  Mpc with  9\%  solar  oxygen
abundance\cite{Pettini04,  Bergvall95}.   Their  metallicities may  be
similar to  that of gas out of  which Population II stars  form in the
early Universe  around redshift from  7 to 12\cite{Wise12}.   An effective
way to estimate the total gas content in extremely metal poor galaxies
is to employ spatially resolved maps of the far-infrared (IR) emitting
dust as tracers of the atomic and molecular gas by multiplying with an
appropriate gas-to-dust ratio (GDR) that is inferred from regions with
little  or no  active  star  formation, a  methodology  that has  been
extensively     demonstrated      in     relatively     metal     rich
galaxies\cite{Leroy11, Sandstrom13}.

\begin{figure*}
\centerline{ \includegraphics[width=0.7\textwidth]{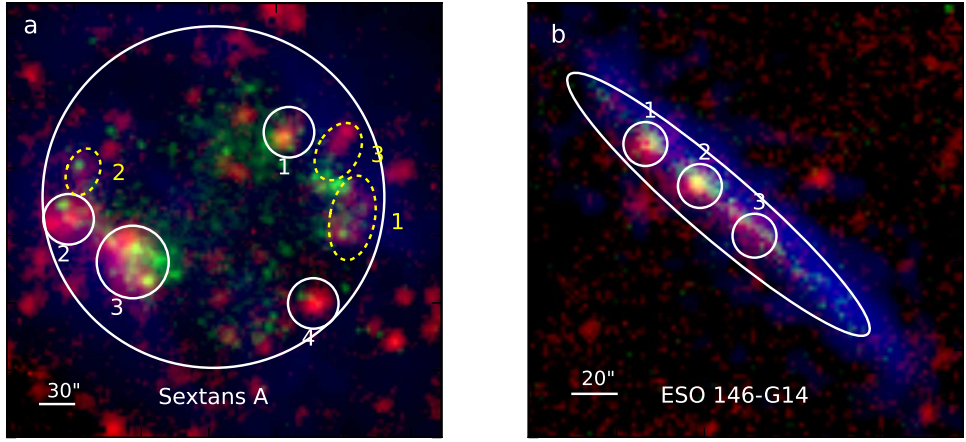} }
\vspace{-4mm}
\caption{  {\bf  False-color, multi-wavelength  images  of our  sample
galaxies}. {\bf  a}, Images of Sextans  A: red, the  sum of {\it Herschel}
160 and 250 $\mu$m data; green,  GALEX far-UV; blue, radio 21 cm data.
The  star-forming disk  is defined  as the  large circle.   The small
circles  indicate individual  dusty star-forming  clumps,  while small
dotted  ellipses  indicate  individual  diffuse regions.   {\bf  b},
Images of  ESO 146-G14: red,  the sum of  Herschel 160 and  250 $\mu$m
data;  green, GALEX  far-UV;  blue, radio  21  cm data.   The disk  is
indicated as  the large  ellipse while individual  star-forming clumps
are shown as small circles.}
\vspace{-4mm}
\end{figure*}

\begin{figure*}
\centerline{\includegraphics[width=1.0\textwidth]{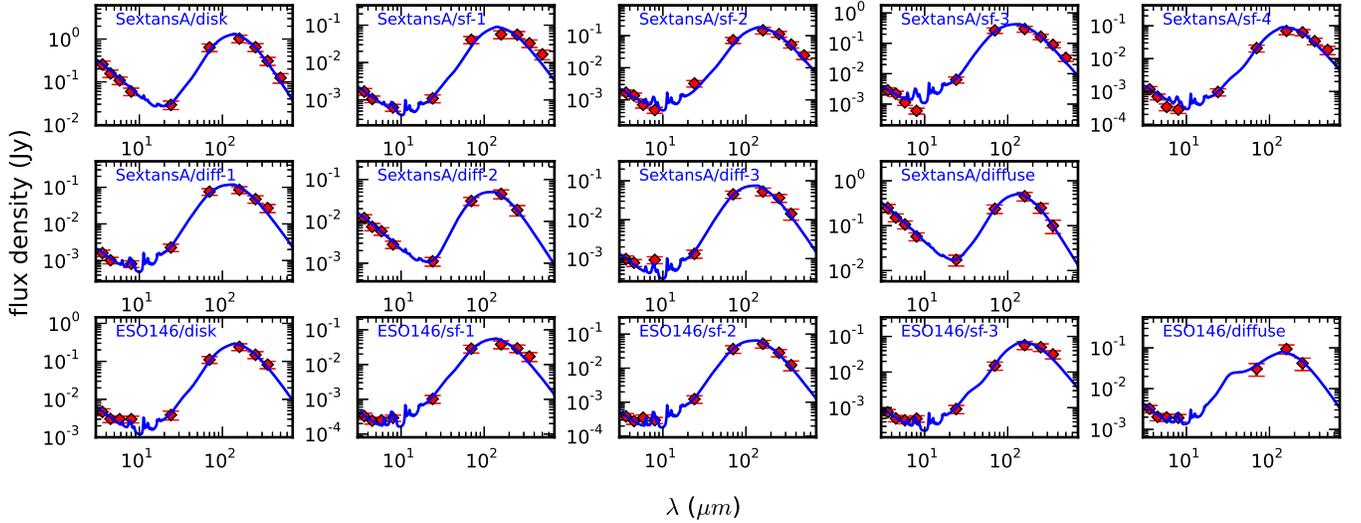} }
\vspace{-4mm}
\caption{{\bf Infrared SEDs of individual 
regions were fitted to derive dust masses:} Red symbols are the {\it Spitzer} and {\it Herschel} photometric points
with 1-$\sigma$ error bars.  
The blue solid line indicates the best-fit by the dust  model\cite{Draine07}.  }
\vspace{-4mm}
\end{figure*}

The infrared observations described in  this paper were carried out at
70, 160, 250,  350 and 500 $\mu$m with  the Photodetector Array Camera
and Spectrometer (PACS)\cite{Poglitsch10} and Spectral and Photometric
Imaging  REceiver (SPIRE)\cite{Griffin10}  onboard the  {\it Herschel}
Space  Observatory.  We complement  our far-IR  data with  mid-IR images
from  the {\it  Spitzer}  Space  Telescope to  construct  the full  IR
spectral energy distributions (SED).  Far-ultraviolet (UV) images from
the {\it GALEX} Space Telescope  archive are used to trace un-obscured
star formation.   Maps of atomic  gas are available in  the literature
for Sextans A\cite{Ott12} and ESO 146-G14\cite{Peters13}.

Figure  1 shows the  multi-wavelength images  of our  sample galaxies.
Based  on the  far-UV image  we defined  the star-forming  disk  as an
ellipse    to   closely    follow   the    10-$\sigma$    ($\sim$   26
AB-mag/arcsec$^{2}$) contour  of the far-UV  emission as shown  in the
Fig. 1 and  listed in Table 1.  Individual  star-forming clumps within
the  disk  are  identified  as  circled  regions  with  elevated  ($>$
3$\sigma$) emission relative to  local disk backgrounds in both far-UV
and 160  $\mu$m bands after  smoothing images to 28  arcsec resolutions.  
The  diffuse emission is measured by  subtracting the total
emission of  all star-forming clumps  in the disk from  the integrated
disk emission.  For Sextans A,  we also identified  several individual
diffuse regions that show extended emission in 70 and 160 $\mu$m bands
but   with  surface   brightness   below  3$\sigma$   of  local   disk
backgrounds. In order to derive  the dust mass, including both diffuse
and  clumped  components, we  fit  the IR  SED  with  a standard  dust
model\cite{Draine07}.   The best-fit SEDs  are shown  in Fig.   2, and
derived dust masses are listed in Table 1.

\begin{table*}
\begin{center}{\bf Table 1: The Properties Of The Sample}\end{center}
\scriptsize
\begin{center}
\begin{tabular}{llllllllllllll}
\hline
\hline
Region     & Right ascension & Declination  &  m$_{a}$$^{*}$,m$_{b}$$^{*}$        & Dust mass     & $M_{\rm HI}/M_{\rm dust}$$\dagger $ &  log$\Sigma_{\rm gas}$$^{\ddagger}$     &   log$\Sigma_{\rm SFR}$$^{\P}$            \\
           &   (J2000)       & (J2000)      &  (arcsec; kpc)                   & (M$_{\odot}$)   &                        & (logM$_{\odot}$/pc$^{2}$)   &  (logM$_{\odot}$/yr/kpc$^{2}$) \\ 
\hline 

SextansA/disk            &10:11:01.4 & -04:41:25   &152x152; 1.06x1.06       &(9.5$^{+1.1}_{-1.0}$)x10$^{3}$      & (5.7$^{+0.6}_{-0.7}$)x10$^{3}$      &                           &                     \\
SextansA/sf-1            &10:10:56.9 & -04:40:27   &22x22; 0.16x0.16         &(9.9$^{+2.5}_{-1.5}$)x10$^{2}$      & (1.3$^{+0.6}_{-0.7}$)x10$^{3}$      & 2.26$^{+0.23}_{-0.22}$    & -2.66$\pm$0.2       \\
SextansA/sf-2            &10:11:10.0 & -04:41:44   &22x22; 0.16x0.16         &(2.$^{+0.2}_{-0.2}$)x10$^{3}$       & (1.3$^{+0.6}_{-0.7}$)x10$^{3}$      & 2.57$^{+0.21}_{-0.21}$    & -2.77$\pm$0.2       \\
SextansA/sf-3            &10:11:06.2 & -04:42:23   &32x32; 0.22x0.22         &(1.8$^{+0.4}_{-0.3}$)x10$^{3}$      & (3.2$^{+0.6}_{-0.7}$)x10$^{3}$      & 2.21$^{+0.23}_{-0.21}$    & -2.32$\pm$0.2       \\
SextansA/sf-4            &10:10:55.5 & -04:42:59   &22x22; 0.16x0.16         &(1.6$^{+0.1}_{-0.1}$)x10$^{3}$      & (4.1$^{+5.8}_{-6.6}$)x10$^{2}$      & 2.46$^{+0.21}_{-0.21}$    & -3.19$\pm$0.2       \\
SextansA/diff-1          &10:10:53.2 & -04:41:43   &38x20; 0.26x0.14         &(5.1$^{+0.8}_{-0.5}$)x10$^{2}$      & (6.9$^{+0.6}_{-0.7}$)x10$^{3}$      &                           &                     \\
SextansA/diff-2          &10:11:09.2 & -04:41:02   &21x14; 0.15x0.10         &(1.8$^{+0.3}_{-0.3}$)x10$^{2}$      & (8.6$^{+0.6}_{-0.7}$)x10$^{3}$      &                           &                     \\
SextansA/diff-3          &10:10:54.0 & -04:40:44   &27x18; 0.19x0.13         &(3.2$^{+0.6}_{-0.3}$)x10$^{2}$      & (6.6$^{+0.6}_{-0.7}$)x10$^{3}$      &                           &                     \\
SextansA/diffuse         &           &             &                         &(3.1$^{+0.3}_{-0.4}$)x10$^{3}$      & (1.4$^{+0.1}_{-0.1}$)x10$^{4}$      &                           &                     \\
ESO146-G14/disk          &22:13:01.3 & -62:04:00   &90x15; 9.34x1.56         &(5.9$^{+0.9}_{-0.5}$)x10$^{5}$      & (2.5$^{+0.2}_{-0.5}$)x10$^{3}$      &                           &                     \\
ESO146-G14/sf-1          &22:13:06.0 & -62:03:33   &10x10; 1.04x1.04         &(7.5$^{+2.1}_{-1.0}$)x10$^{4}$      & (1.6$^{+0.2}_{-0.5}$)x10$^{3}$      & 1.21$^{+0.24}_{-0.22}$    & -3.46$\pm$0.2       \\
ESO146-G14/sf-2          &22:13:02.5 & -62:03:52   &10x10; 1.04x1.04         &(6.2$^{+0.9}_{-0.8}$)x10$^{4}$      & (2.7$^{+0.2}_{-0.5}$)x10$^{3}$      & 1.12$^{+0.22}_{-0.22}$    & -3.26$\pm$0.2       \\
ESO146-G14/sf-3          &22:12:59.0 & -62:04:14   &10x10; 1.04x1.04         &(2.7$^{+0.2}_{-0.2}$)x10$^{5}$      & (5.$^{+2.2}_{-4.7}$)x10$^{2}$       & 1.77$^{+0.21}_{-0.21}$    & -3.65$\pm$0.2       \\
ESO146-G14/diffuse       &           &             &                         &(2.5$^{+0.3}_{-0.3}$)x10$^{5}$      & (4.4$^{+0.2}_{-0.5}$)x10$^{3}$      &                           &                     \\
\hline
\hline
\end{tabular}
\end{center}
$^{*}$Major and minor axis lengths are given in arcsec and kpc.

$^{\dagger}$The atomic gas to dust mass ratio. The atomic gas includes Helium by multiplying
HI gas by a factor of 1.36.

$^{\ddagger}$ Surface densities of total gas masses for star-forming clumps are derived from their dust masses 
multiplied with gas-to-dust ratio of the integrated diffuse emission, with the inclination correction based 
on the defined disk ellipse.

$^{\P}$ Surface densities of SFRs are derived from the combination of IR and far-UV tracers\cite{Leroy08}, with the inclination corrected based on the defined disk ellipse.
\end{table*}

With  spatially resolved dust  and HI  maps, the  total gas  masses of
individual  star-forming clumps  can be  derived by  multiplying their
dust masses with  the appropriate GDR based on  regions with little or
no    star   formation.     As    is   usually    done   for    nearby
galaxies\cite{Leroy11, Sandstrom13} the GDR in the star-forming clumps
is taken as the  ratio of atomic gas to dust in  the diffuse region of
the disk.  This  works because (1) the atomic  gas dominates the total
gas mass in  the diffuse regions, and (2) the  GDR is roughly constant
in    star-forming    disks    after    removing    the    metallicity
gradients\cite{Sandstrom13, Draine13}.  Dwarf galaxies in general show
little     or     no     metallicity    gradients     across     their
disks\cite{Westmoquette13}, including Sextans A, which has a variation
of less than  0.1 dex\cite{Kniazev05}.  Table 1 lists  the derived gas
masses  of individual  star-forming clumps  corrected  for inclination
based   on    the   GDR    of   the   integrated    diffuse   emission
(GDR=1.4$\times$10$^{4}$  for Sextans  A and  4,400 for  ESO 146-G14).
For  Sextans A, three  individual diffuse  regions have  similar GDR's
that are  only a factor of 1.5-2  lower than the one  derived from the
integrated diffuse  emission.  Adopting  the GDR of  140 at  the solar
metallicity\cite{Draine07}, our derived GDR  of diffuse regions of two
galaxies scales  roughly with the metallicity  as $1/Z^{1.5-1.7}$. For
each star-forming clump, the star formation rate (SFR) is estimated by
combining  the FUV-based  (unobscured)  and 24$\mu$m-based  (obscured)
SFRs\cite{Leroy08}.  The  uncertainties in the derived  gas masses and
SFRs are estimated to be around 0.3 dex and 0.2 dex, respectively.

Figure 3 shows the distribution  of seven dusty star-forming clumps in
the  plane  of SFR  surface  densities  vs.   total gas  mass  surface
densities,  compared to  spirals  and merging  galaxies\cite{Bigiel08,
Daddi10}. In sharp  contrast to deriving the gas  surface density from
the HI gas alone (open symbols  in the figure), when using the dust to
estimate the  total gas, the metal-poor star-forming  clumps appear to
have significantly  lower star formation  efficiencies (SFEs) compared
to those found in metal-rich galaxies, or those derived for the clumps
using the HI gas alone.  Four extremely metal poor clumps in Sextans A
show almost  two orders  of magnitude lower  SFEs compared  to spirals
when  measured over the  similar physical  scales.  This  result still
holds if  we adopt a GDR  of individual diffuse  regions, which causes
the  gas densities only  drop by  0.2-0.3 dex.   For ESO  146-G14, one
star-forming  clump  shows  significantly  (100) lower  SFEs  and  the
remaining two  have SFEs about  a factor of  10 lower than  spirals at
sub-kpc scales and  similar gas densities.  If any  dark molecular gas
is  present  in the  diffuse  region, the derived  SFEs  would be  even
lower.  For our  seven metal  poor  clumps as  a group,  the K-S  test
indicates a  probability of  only 10$^{-4}$ that  their SFRs  have the
same distribution  as the  SFRs of spiral  galaxies at  comparable gas
densities.

As  illustrated in  Fig.   3,  the derived  dust-based  gas masses  of
individual  clumps  are  much  higher  than  the  atomic  gas  masses,
indicating  high   molecular  gas  fractions.   By
subtracting the  observed atomic gas  from the derived  dust-based gas
mass  for our  seven star-forming  clumps,  we find  that the  derived
molecular gas mass is on average $\sim$ 6 times larger than the atomic
gas  mass.   For the  star-forming  clumps,  the  median and  standard
deviation     of     the     molecular    SFE     is     log(SFE$_{\rm
H_{2}}$[yr$^{-1}$]))=-10.8$\pm$0.6.   The corresponding  molecular gas
depletion   time   in    logarithm   is   log($\tau^{\rm   H_{2}}_{\rm
dep}$[Gyr])=1.8$\pm$0.6.   This  is   much  larger  than  the  typical
depletion  time at  sub-kpc scales  of  local spirals  or dwarfs  with
oxygen  abundance  above  20\%  solar which  is  about  log($\tau^{\rm
H_{2}}_{\rm dep}$[Gyr])=0.3\cite{Bigiel08, Bolatto11}.  Therefore star
formation  is strongly  suppressed  in the  clumps,  whether the  star
formation is scaled by  total gas mass or by molecular gas
mass alone.

Although models\cite{Ostriker10, Bolatto11, Krumholz13} of metallicity
regulated  star formation  predict significantly  reduced SFEs  at our
metallicities  as illustrated in  Fig.  3,  the gas  in the  models is
nevertheless mainly  atomic, in contrast  with our findings.   The low
SFE in the model is caused by a greatly reduced formation of molecular
gas on the  surface of dust grains \cite{Krumholz13}  or possibly tied
to enhanced heating of the atomic gas\cite{Bolatto11}.  On the other hand,
if we increase the depletion time of molecular gas in the models by an
order of  magnitude to $\sim$  20 Gyr, the  gas in the model  is still
mostly atomic at  our observed gas densities.  Studies\cite{Cormier14}
of HI dominated (by assuming low molecular gas fractions) dwarf
galaxies  suggest that  their SFEs  are not  significantly  lower than
spiral  galaxies.    Our  results   extend  these previous findings   to  lower
metallicity,  and suggest  high molecular  gas fractions  do  exist in
star-forming clumps of extremely low metallicity.

The nature of  this excess dust-based gas mass  is still unclear. This
gas is  most likely in  the molecular phase  as the cold HI  should be
detected by  21 cm observations.  If  it is cold  molecular gas, there
should be  associated CO  emission.  Extended Data  Table 6  lists the
predicted CO flux for these metal poor star-forming clumps by assuming
the CO-to-H$_{2}$ factor,  $\alpha_{\rm CO}$=500 $M_{\odot}$ pc$^{-2}$
(K km s$^{-1}$)$^{-1}$,  which is appropriate for regions  of such low
metallicity.   It  should  be  noted  however,  that  there  is  large
uncertainty in this value\cite{Bolatto13}.  Region "sf-3" in Sextans A
does indeed have  mm-wave observations\cite{Taylor98}.  Accounting for
a filling  factor of  one third of  dust emission  in the CO  beam, we
estimate  a 3-$\sigma$  upper limit  to the  CO flux  that is  about a
factor of  three above our  predicted value. Therefore much  deeper mm
observations would be needed to detect the CO emission from the excess
cold  molecular gas  in  Sextans  A.  It  is  unclear what  mechanisms
prevents this  abundant molecular gas  from forming new stars.   It is
possible  that the  molecular gas  does  not effectively  cool due  to
intense  radiation fields,  slowing  the SFRs  in these  environments.
Warm   H$_{2}$   gas   with   surface   densities  as   high   as   50
M$_{\odot}$/pc$^{2}$     is    seen     in    some     blue    compact
dwarfs\cite{Hunt10}. Although  our two  galaxies are not  blue compact
dwarfs, the SFR  surface densities of the star-forming  regions in our
galaxies are comparable  to those found in blue  compact dwarfs.  This
similarity suggests the possible  presence of abundant warm H$_{2}$ in
our two extremely metal poor galaxies.  The Extended Data Table 6 also lists
the  predicted H$_{2}$  $S$(1) 17.03  $\mu$m  line flux  based on  the
example  of Mrk  996\cite{Hunt10}.   There is  archived {\it  Spitzer}
spectroscopic observations of the  region ``sf-3'' of Sextans A. Based
on  the archived  reduced data,  after accounting  for  the difference
between the {\it  Spitzer} aperture and the size  of our ``sf-3'', the
observed H$_{2}$ 17.03 $\mu$m  flux is about 4x10$^{-16}$ W/m$^{2}$, a
factor of two lower than our predicted value.

The extremely metal  poor galaxies may provide a  close-up view of the
highly inefficient  star formation occurring in galaxies  in the early
Universe where Population  II stars form out of  gas whose metallicity
is 1/10 Solar or  less\cite{Wise12}.  The suppressed SFEs in extremely
low metallicity galaxies at early epochs may be able to reconcile some
tensions between observations and  theoretical models for early galaxy
evolution\cite{Kuhlen12}.

\begin{figure}
\centerline{\includegraphics[width=0.5\textwidth]{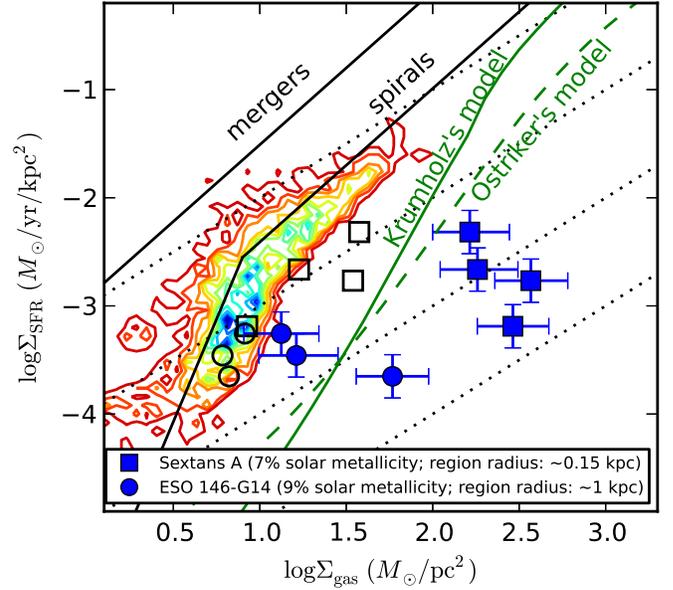} }
\vspace{-4mm}
\caption{ {\bf Seven metal poor star-forming clumps show extremely low
star formation efficiencies:} Our  data with 1-$\sigma$ errors (filled
symbols for  dust-based total gas  mass while open symbols  for atomic
gas   only),    were   compared    to   spiral   disks    at   sub-kpc
scales\cite{Bigiel08}  (color  contour),  and  integrated  spiral  and
mergers\cite{Daddi10} (black lines)  in the plane of the  SFR vs.  gas
surface  densities.   The  green  solid  and  dashed  lines  represent
predictions of the model\cite{Krumholz13} at 8\% solar metallicity and
a  clumping  factor  of  1,  and model\cite{Bolatto11}  at  8\%  solar
metallicity.  Dotted lines indicate constant  SFEs of, from the top to
the bottom, 10$^{-9, -10, -11, -12}$ yr$^{-1}$.}
\vspace{-4mm}
\end{figure}

\begin{addendum}
\item  [Acknowledgements]  

Y.S.   acknowledges  support  for   this  work  from  Natural  Science
Foundation of China (NSFC) under  the grant of 11373021, the Strategic
Priority Research Program ``The Emergence of Cosmological Structures''
of the Chinese  Academy of Sciences (CAS), Grant  No.  XDB09000000, and
Nanjing University  985 grant.   Y.G.  acknowledges support  from NSFC
grants   (11173059  and   11390373)   and  CAS   Program  (Grant   No.
XDB09000000).    J.W.    is   supported   by  National   973   program
(2012CB821805) and NSFC  grant (11173013).  Z.Z.  acknowledges support
from  the European  Research Council  (ERC)  in the  form of  Advanced
Grant,  {\sc cosmicism}.   Q.G.  is  supported under  the  NSFC grants
(11273015 and  11133001) and the National  973 program (2013CB834905).
We are grateful  to F.  Bigiel for making  their data points available
to plot contours in Figure 3,  S.  P.  C.  Peters for making available
their HI gas map of ESO 146-G14 to us, and L.  Piazzo for help in {\it
Herschel} data  reduction. Herschel is  an ESA space  observatory with
science  instruments provided  by European-led  Principal Investigator
consortia and  with important participation from NASA.   This work was
supported in part by the Spitzer Space Telescope, which is operated by
the  Jet  Propulsion Laboratory,  California  Institute of  Technology
under a contract  with NASA.  It was also supported in  part by a NASA
Herschel grant (OT2$\_$yshi$\_$3) issued by JPL/Caltech.

\item[Author  Contributions] 

Y.S. led the Herschel proposal, Herschel data reduction and writing of
the manuscript.   L.A. help develop Herschel  observations and writing
of  the manuscript.   G.H.  and  S.S. assisted  in the Herschel proposal.
  All  authors discussed  and commented on the manuscript.

\item[Author  Information]  

Reprints    and    permissions    information    is    available    at
www.nature.com/reprints.  The authors  declare no  competing financial
interests.  Readers  are welcome to  comment on the online  version of
the  paper.   Correspondence  and  requests for  materials  should  be
addressed to YS (yshipku@gmail.com).

\end{addendum}

\clearpage


\include{method}

\end{document}

%% file: method.tex
\setcounter{page}{1}
\setcounter{figure}{0}
\setcounter{table}{0}
\renewcommand{\thefigure}{\arabic{figure}}
\renewcommand{\thetable}{\arabic{table}}
\renewcommand{\figurename}{Extended Data Figure}
\renewcommand{\tablename}{Extended Data Table}

\begin{center}
{\bf \Large \uppercase{Methods} }
\end{center}

\section{Observations and data reduction} 

{\it Herschel}  infrared images were generated through  scan map modes
at 70 and 160  $\mu$m with PACS, and small map modes  at 250, 350,
and 500 $\mu$m  with SPIRE (PI: Y.  Shi,  PID: OT2$\_$yshi$\_$3).  The
half power beam  widths at these wavelengths are  about 5$''$, 12$''$,
20$''$, 28$''$  and 39$''$, respectively.   The mapping field  size at
each band is set to be at least 1.5 times the optical size $D_{25}$ of
the  galaxy  where  $D_{25}$  is  the  $B$-band  isophote  at  25  mag
arcsec$^{-2}$,  which is  large enough  to provide  blank sky  for sky
removal.  Excluding overheads, the effective integration times per sky
position in two PACS bands are 1.9 hr and 1.6 hr for Sextans A and ESO
146-G14, respectively, and  in three SPIRE bands is 6  min for the two
targets.   For Sextans  A  at 160  $\mu$m,  additional {\it  Herschel}
archived data (PI: D.  Hunter) with similar exposure time to ours were
further combined  with our own  observations.  The data  reduction was
performed with  unimap version 5.5  \cite{Traficante11, Piazzo12} with
procedures basically  following the  standard one.  To  better recover
the extended emission, the GLS map  maker starts with the zero map and
the length of median filter for  the PGLS algorithm is set to be twice
the default value (=60).  Unimap is a Matlab-based {\it Herschel} data
processing software.  Unimap takes as  input the level 1 pipeline data
as  produced by  {\it  Herschel} Standard  Product Generator  (Version
11.1.0 for this  work) and identifies signals in  time ordered pixels.
After removing glitch  and drift, the final maps  were made with pixel
scales of  1$''$, 2$''$, 4$''$, 6$''$  and 8$''$ at 70,  160, 250, 350
and 500 $\mu$m, respectively.

The reduced  far-UV images were  obtained from the GALEX  data archive
hosted  by the Multi-Mission  archive at  the Space  Telescope Science
Institute.   GALEX  has  a  spatial  resolution  about  5$''$  in  the
far-UV\cite{Morrissey07}.   The  exposure  times  for  Sextans  A  and
ESO146-G14  are 1698  sec and  111 sec,  respectively.  Sextans  A has
reduced {\it  Spitzer} images at  3.6, 4.5, 5.8,  8.0, 24, 70  and 160
$\mu$m obtained by the  LVL program\cite{Dale09} available in the NASA
infrared science  archive.  The corresponding  spatial resolutions are
about  5$''$$\times$$\lambda$/24$\mu$m.  For  ESO146-G14, the  data at
3.6, 4.5, 5.8,  8.0 and 24 $\mu$m were available  in the {\it Spitzer}
Heritage Archive at the {\it  Spitzer} Science Center and the archived
enhanced  imaging products  were used.  The {\it  Spitzer} 70  and 160
$\mu$m integrated fluxes\cite{Engelbracht08}  of ESO 146-G14 were used
to compared to our PACS measurements.

\section{Infrared flux measurements}

Our {\it  Herschel} flux measurements start  with aperture definitions
followed by sky  subtractions.  As shown in Extended  Data Fig. 1, the
aperture  of the star-forming  disk of  each galaxy  is defined  as an
ellipse to closely follow the 10 $\sigma$ contour above the sky of the
far-UV image,  corresponding to surface brightness levels  of 25.9 and
26.2 mag arcsec$^{-2}$  (AB magnitude) for Sextans A  and ESO 146-G14,
respectively. The results of this study change little if 5-$\sigma$ or
20-$\sigma$ contours  were used to  define the disk  aperture.  Within
each  star-forming disk,  star-forming  clumps are  defined as  circle
regions showing both elevated  ($>$ 3-$\sigma$) far-UV and IR emission
at 160 $\mu$m  after convolving two images to  resolutions at PACS 350
$\mu$m (28  arcsec). Here  the $\sigma$ is  the standard  deviation of
pixel values within the star-forming disk.  The clump radius is listed
in  Extended Data Table  1. For  Sextans A,  we also  identified three
individual diffuse regions that are below 3-$\sigma$ local backgrounds
but show extended  emission at 70 and 160  $\mu$m resolutions.  Within
the  disk, IR  point sources  that  do not  have corresponding  far-UV
counterparts  are   identified  as  background   sources  rather  than
star-forming regions in the disk,  since none of the identified clumps
has IR/far-UV  flux ratios  smaller than about  0.2. 

The sky annuli were defined to  be between 1.1 times and 1.5 times the
disk aperture.  The  mode of the sky pixel  brightness distribution is
subtracted  from  the  image  in  each case.   However,  since  faint,
undetected  background   sources  can  make   the  noise  distribution
non-Gaussian, we also test the  validity of our results by subtracting
off the mean value of the  sky after masking out bright sources in the
sky annuli.

In total, we  derive three types of flux  measurements for each region
as listed in Extended Data Table 1.  For the first, referred to as m1,
we use the mode of the  sky brightness after masking off all potential
background sources in the disk. For  the second, referred to at m2, we
again subtract  off the  mode of the  sky distribution, but  treat the
suspected background sources as embedded star-formation regions in the
disk.   For the  third,  referred to  as  m3, we  first  mask out  all
potential  background  sources, then  subtract  the  mean  of the  sky
pixels.  All  potential background  sources and bright  sources within
the sky annuli  are identified through sextractor\cite{Bertin96} with
further visual  checks. We  use the first  flux estimates, m1,  as the
fiducial for the analysis presented here.

For images at each wavelength, aperture photometry was performed after
subtracting the sky background.  The aperture corrections were further
applied  at each wavelength  based on  the corresponding  point spread
function  at that  wavelength.  Diffuse  emission within  the  disk is
measured  as  the  residual   after  subtracting  the  flux  from  all
identified star-forming  clumps.  The result is  presented in Extended
Data Table 1 where each  region has three measurements detailed above,
referred as m1, m2 and m3 measurements.

The flux measurements in the {\it  Spitzer} band were carried out in a
similar way to the {\it Herschel} m-1 method and the result is listed in Extended Data Table 2.

\section{Infrared flux uncertainty  estimates}

The {\it Herschel} flux uncertainties are given by the following formula
\begin{equation}
\sigma = (\sigma_{\rm photon, confusion}^{2}{\times}A_{\rm region}+\frac{\sigma_{\rm photon, confusion}^2}{A_{\rm sky}} A_{\rm region}^{2}+
\sigma_{\rm PSF-offset}^{2}) + \sigma_{\rm abs-calibration}^{2}
\end{equation} , where $A_{\rm region}$ and $A_{\rm sky}$ are the area
of target regions and  sky annuli, respectively.  $\sigma_{\rm photon,
confusion}$ is  the scatter of the sky  pixel brightness distribution.
Extended  Data  Table 3  compares  our  measured $\sigma_{\rm  photon,
confusion}$  to the  predicted photon  and confusion  noises estimated
using the {\it  Herschel} Observing Tool (HSPOT) for  our targets. The
noise in our images is consistent  with the quadratic sum of the HSPOT
photon and confusion noise to within a factor of two.  The second term
in  the  above   equation  gives  the  scatter  of   the  derived  sky
brightness. $\sigma_{\rm  PSF-offset}$ is the  flux uncertainty caused
by the accuracy  in positioning an aperture onto  a given star-forming
clump.  For  each star-forming region,  we estimated this  by randomly
offsetting the peak of a modelled  PSF to 1/2 the Nyquist sampled beam
and  measuring the flux  variation within  the given  source aperture.
The final term is the absolute flux calibration error taken to be 20\%
across  all  wavelengths  based  on  the  PACS  and  SPIRE  instrument
handbooks  as well  as systematic  comparisonss\cite{Sauvage11, Ali11,
Paladini12, Paladini13}  between PACS and  MIPS measurement.  Extended
Data Table 1 listed the quadratic  sum of the first three errors.  The
final error term is added in quadrature when doing the SED fitting but
not used in Extended  Data Table 1 as it is a  systematic error of the
{\it  Herschel}  Space  Telescope.   Our estimated  errors  are  quite
reasonable  compared to  the expected  point source  flux  errors from
HSPOT (also  listed in Extended Table  3).  Note that  although in the
SPIRE bands  the confusion noise is  2-3 times higher  than the photon
noise, this can  be mitigated by using a PACS  160-micron prior on the
position\cite{Elbaz11}.  We can further compare our noise estimates to
those  from other {\it  Herschel} observations  of similar  depth. For
example,  the {\it Herschel}  lensing survey\cite{Egami10}  reported a
1-$\sigma$ point-source depth of 2.4 mJy  at 250 $\mu$m and 3.4 mJy at
350 $\mu$m  for on-source  exposure per sky  position of 36  mins with
position priors from short wavelengths, compared to our 2-4 mJy at 250
$\mu$m  and  3-5  mJy at  250  $\mu$m  for  our  6-10 mins  on  source
exposures.

We  further carried  out additional  checks  on the  measured flux  by
comparing {\it Spitzer} and  {\it Herschel} photometry. For Sextans A,
we found  that individual star-forming  clumps as well as  the diffuse
region  have 70  $\mu$m  {\it Herschel}  fluxes  consistent with  {\it
Spitzer} measurements within 30\%. And the integrated light of Sextans
A and ESO146-G14 at both 70  and 160 $\mu$m are also consistent within
30\% between the {\it Spitzer} and {\it Herschel} data-sets.

To  check the  possibility that  the diffuse  emission is  due  to the
background  fluctuations  or not.   We  randomly  position the  source
aperture  over the  observed  field  of view,  and  then compared  the
measured   fluxes  to  the   quoted  error   of  the   target  diffuse
emission. For ESO 146-G14, we can randomly position about 30 apertures
and found that none of them have  S/N larger than 3 at bands where the
diffuse emission  is detected.  For Sextans A,  the observed  field of
view is not large enough for us to perform similar exercises.

\section{IR SED fitting and dust mass measurements}

We fit the infrared data  with the dust models\cite{Draine07} in order
to estimate  the dust mass  of each region.   As shown above,  we have
three  types of flux  measurements, and  fit all  three with  the dust
model.  We  choose a Milky Way  grain size distribution\cite{Draine07}
and fix the  PAH fraction to the minimum (the total  dust mass that is
in PAHs  $q_{\rm PAH}$=0.47\%) given  the low metallicity  (the result
does not change if this parameter  is set free).  To further check the
effect  of  different  dust  grains,   for  the  first  type  of  flux
measurements (m1), SMC  and LMC dust grains that  have different grain
compositions and size distributions are also explored. Overall we thus
have five  dust mass measurements for  each region, three  of them are
for  different  types  of  flux  measurements with  Milky  Way  grains
(referred as  m1-MW, m2-MW,  m3-MW), two are  for two  different grain
size  distributions fitted  to  the first  type  of flux  measurements
(m1-SMC and m1-LMC).

In the  following we take  the m1-MW as  an example to  illustrate the
fitting procedure.   The results are plotted  in Fig. 2  and listed in
Extended Data  Table 4.  To do  the fit, a 4000  K black-body spectrum
was first  added to represent  the emission from  stellar photospheres
which dominates at  $<$ 10 $\mu$m.  The model was  then left with four
free parameters, including the  dust mass, the minimum ($U_{\rm min}$)
and maximum  intensity ($U_{\rm max}$) of the  stellar radiation field
that is responsible for heating the dust and the fraction (1-$\gamma$)
of  dust exposed  to the  minimum starlight  intensity  (i.e.  $U_{\rm
min}$).    Similar   to   studies   of  dust   emission   in   spirals
\cite{Sandstrom13}, $U_{\rm max}$ was  further fixed to the maximum of
10$^{6}$.  We  then performed SED  fitting with three  free parameters
for Sextans A and ESO 146-G14.

As listed  in Extended  Data Table 4,  the reduced $\chi^{2}$  for the
majority of the fits have values of around unity, while sf-1, sf-2 and
diff-3 of  Sextans-A have reduced  $\chi^{2}$ around 10. As shown in
Fig.  2, the  160 $\mu$m  photometry of  sf-1 and  diff-3  shows large
deviations from  the best  fit, while 24  and 70 $\mu$m  photometry of
sf-b deviates largely from the  best fit.  Uncertainties in the derived
dust mass were  estimated by performing 100 fits  to each source after
adding in Gaussian noise.

We also carried  out simple modified black-body fitting  to IR SEDs of
Sextans A  and ESO  146-G14 that have  enough far-IR  photometric data
points. As  listed in Extended Data  Table 4, the  dust temperature of
star-forming  clumps is  between 30  and 50  K while  the dust  in the
diffuse region is about 30-40 K.

\section{Measurements of SFR and gas mass surface densities}

The SFRs of star-forming clumps were measured by combining the FUV and
24-micron   data,  which   represent  the   unobscured   and  obscured
star-formation,  respectively\cite{Leroy08}.   The  derived  SFRs  are
uniformly  assigned a  0.2 dex  error  to account  for the  systematic
errors in deriving SFRs from UV and IR photometry (the photon noise is
comparatively small).   The SFR  surface density is  further corrected
for the  inclination based on  the defined disk ellipsity.   The final
result is listed in Table 1.

With derived dust  masses, we estimated the GDR  of the diffuse region
as  the ratio  of atomic  to  dust mass.   The GDR  of the  integrated
diffuse region  is then applied  to individual star-forming  clumps to
derive  the total  gas  mass and  thus  the gas  mass surface  density
($\Sigma_{\rm gas}$).  Extended Data Table 5 lists the result for five
fits that are m1-MW, m2-MW,  m3-MW, m1-SMC and m1-LMC.  The associated
uncertainties of $\Sigma_{\rm gas}$ are the quadratic sum of errors of
dust mass measurements, errors  of GDRs of diffuse regions contributed
by uncertainties on HI and dust mass estimates of diffuse regions, 0.2
dex for the  GDR variation across the disk based  on studies of spiral
galaxies\cite{Sandstrom13}.  The result of the  m1-MW fit is used as a
fiducial, as listed  in Table 1 and shown in  Fig.  3. Our conclusions
of significantly reduced SFEs  in seven metal poor star-forming clumps
change little if adopting other fitting results in Extended Data Table
5.  In addition,  there are some concerns that the  PACS may miss some
extended emission,  although this  is not seen  by our  comparisons to
{\it Spitzer}  and in investigations  by others\cite{Sauvage11, Ali11,
Paladini12,  Paladini13}.  To  test this  effect, we artifically increased 
the  PACS  fluxes of the diffuse emission  by 30\%  while keeping
the SPIRE fluxes as they  were, the resulting surface densities of gas
masses of star-forming clumps drop  by no more than  0.1 dex. In additiona,
three individual diffuse regions of  Sextans A has similar GDR, only a
factor of  1.5-2 lower than  that of the integrated  diffuse emission,
indicating that our GDR estimate is reasonable.

We investigate if the  derived $\Sigma_{\rm gas}$ can be significantly
lowered  by forcing  changes  in dust  model parameters,  specifically
raising $U_{\rm min}$ which can  result in lower dust masses and hence
lower gas  surface densities  and higher star  formation efficiencies.
In the following discussion, we take  the m1-MW fit as the case study.
For  both targets,  the  best-fit  $U_{\rm min}$  of  all regions  are
relatively  small. We  thus keep  the best-fit  $U_{\rm min}$  for the
diffuse region  but gradually  increase $U_{\rm min}$  of star-forming
clumps  to  decrease  their  $\Sigma_{\rm  gas}$.  We  find  that  the
star-forming clumps  in Sextans A  can move into  the spiral galaxy
regime of Fig. 3 if the U$_{\rm min}$ rises above 20.  However, in this
case the  corresponding chi-squared rises to  40-60.  For star-forming
clumps in ESO 146-G14, the $U_{\rm min}$ needs to be larger than 15 to
move into  the spiral regime however  these fits are  again poor, with
chi-squared values of 10-30.  Therefore the significantly reduced SFEs
of star-forming clumps  in Sextans A and ESO  146-G14 should be robust
to the change in their $U_{\rm min}$.


\setcounter{figure}{0}

\begin{table*}
\footnotesize
\begin{center}
\begin{tabular}{llllllllllllll}
\hline
region                   &  Right ascension & Declination   &  sizes(ma,mb)     &  f(70$\mu$m)  & f(160$\mu$m) & f(250$\mu$m) & f(350$\mu$m) & f(500$\mu$m) \\
                         &   (J2000)        &  (J2000)      &   (arcsec)        &  (mJy)        & (mJy)        & (mJy)        & (mJy)        & (mJy)\\ 
\hline 

SextansA/disk            & 10 11 01.4 & -04 41 25   &152.0x152.0              &  636$\pm$  16           & 1024$\pm$  18           &  644$\pm$  30           &  308$\pm$  23           &  124$\pm$  18           \\
                         &            &             &                         &  658$\pm$  16           & 1098$\pm$  18           &  722$\pm$  30           &  356$\pm$  23           &  155$\pm$  18           \\
                         &            &             &                         &  605$\pm$  16           &  979$\pm$  18           &  557$\pm$  30           &  236$\pm$  23           &   78$\pm$  18           \\
SextansA/sf-1            & 10 10 56.9 & -04 40 27   &22.5x22.5                &   40$\pm$   2           &   56$\pm$   7           &   55$\pm$   3           &   32$\pm$   3           &   16$\pm$   3           \\
                         &            &             &                         &   40$\pm$   2           &   56$\pm$   7           &   55$\pm$   3           &   32$\pm$   3           &   16$\pm$   3           \\
                         &            &             &                         &   39$\pm$   2           &   55$\pm$   7           &   53$\pm$   3           &   30$\pm$   3           &   14$\pm$   3           \\
SextansA/sf-2            & 10 11 10.0 & -04 41 44   &22.5x22.5                &   72$\pm$   3           &  147$\pm$  18           &  111$\pm$   4           &   52$\pm$   4           &   24$\pm$   3           \\
                         &            &             &                         &   72$\pm$   3           &  147$\pm$  18           &  111$\pm$   4           &   52$\pm$   4           &   24$\pm$   3           \\
                         &            &             &                         &   71$\pm$   3           &  146$\pm$  18           &  109$\pm$   4           &   50$\pm$   4           &   22$\pm$   3           \\
SextansA/sf-3            & 10 11 06.2 & -04 42 23   &32.3x32.0                &  265$\pm$   4           &  296$\pm$  24           &  164$\pm$   5           &   89$\pm$   4           &   33$\pm$   3           \\
                         &            &             &                         &  265$\pm$   4           &  296$\pm$  24           &  164$\pm$   5           &   89$\pm$   4           &   33$\pm$   3           \\
                         &            &             &                         &  264$\pm$   4           &  294$\pm$  24           &  160$\pm$   5           &   85$\pm$   4           &   30$\pm$   3           \\
SextansA/sf-4            & 10 10 55.5 & -04 42 59   &22.5x22.5                &   20$\pm$   2           &   69$\pm$   8           &   62$\pm$   3           &   34$\pm$   3           &   18$\pm$   3           \\
                         &            &             &                         &   20$\pm$   2           &   69$\pm$   8           &   62$\pm$   3           &   34$\pm$   3           &   18$\pm$   3           \\
                         &            &             &                         &   20$\pm$   2           &   68$\pm$   8           &   60$\pm$   3           &   31$\pm$   3           &   16$\pm$   3           \\
SextansA/diff-1          & 10 10 53.2 & -04 41 43   &38.0x20.0                &   75$\pm$   3           &   85$\pm$   5           &   47$\pm$   4           &   27$\pm$   3           &   $<$13                 \\
                         &            &             &                         &   75$\pm$   3           &   85$\pm$   5           &   47$\pm$   4           &   27$\pm$   3           &   $<$13                 \\
                         &            &             &                         &   74$\pm$   3           &   83$\pm$   5           &   43$\pm$   4           &   23$\pm$   3           &   $<$13                 \\
SextansA/diff-2          & 10 11 09.2 & -04 41 02   &21.4x14.6                &   30$\pm$   2           &   45$\pm$   6           &   18$\pm$   3           &   $<$10                 &   $<$12                 \\
                         &            &             &                         &   30$\pm$   2           &   45$\pm$   6           &   18$\pm$   3           &   $<$10                 &   $<$12                 \\
                         &            &             &                         &   30$\pm$   2           &   44$\pm$   5           &   16$\pm$   3           &   $<$10                 &   $<$12                 \\
SextansA/diff-3          & 10 10 54.0 & -04 40 44   &27.5x18.5                &   44$\pm$   2           &   52$\pm$   5           &   36$\pm$   4           &   14$\pm$   3           &   $<$13                 \\
                         &            &             &                         &   44$\pm$   2           &   52$\pm$   5           &   36$\pm$   4           &   14$\pm$   3           &   $<$13                 \\
                         &            &             &                         &   44$\pm$   2           &   51$\pm$   5           &   33$\pm$   4           &   11$\pm$   3           &   $<$13                 \\
SextansA/diffuse         &            &             &                         &  237$\pm$  18           &  453$\pm$  39           &  248$\pm$  33           &   99$\pm$  26           &   $<$69                 \\
                         &            &             &                         &  258$\pm$  18           &  527$\pm$  39           &  326$\pm$  33           &  147$\pm$  26           &   $<$69                 \\
                         &            &             &                         &  210$\pm$  18           &  414$\pm$  39           &  173$\pm$  33           &   $<$78                 &   $<$69                 \\
ESO146-G14/disk          & 22 13 01.3 & -62 04 00   &90.0x15.0                &  110$\pm$   3           &  241$\pm$   5           &  148$\pm$   7           &   81$\pm$   7           &                  \\
                         &            &             &                         &  110$\pm$   3           &  241$\pm$   5           &  148$\pm$   7           &   81$\pm$   7           &                  \\
                         &            &             &                         &  110$\pm$   3           &  238$\pm$   5           &  142$\pm$   7           &   81$\pm$   7           &                  \\
ESO146-G14/sf-1          & 22 13 06.0 & -62 03 33   &10.0x10.0                &   28$\pm$   4           &   38$\pm$   6           &   29$\pm$   4           &   17$\pm$   3           &                  \\
                         &            &             &                         &   28$\pm$   4           &   38$\pm$   6           &   29$\pm$   4           &   17$\pm$   3           &                  \\
                         &            &             &                         &   28$\pm$   4           &   37$\pm$   6           &   28$\pm$   3           &   16$\pm$   3           &                  \\
ESO146-G14/sf-2          & 22 13 02.5 & -62 03 52   &10.0x10.0                &   36$\pm$   5           &   52$\pm$   8           &   28$\pm$   3           &   12$\pm$   3           &                  \\
                         &            &             &                         &   36$\pm$   5           &   52$\pm$   8           &   28$\pm$   3           &   12$\pm$   3           &                  \\
                         &            &             &                         &   36$\pm$   5           &   51$\pm$   8           &   27$\pm$   3           &   12$\pm$   3           &                  \\
ESO146-G14/sf-3          & 22 12 59.0 & -62 04 14   &10.0x10.0                &   15$\pm$   2           &   57$\pm$   9           &   49$\pm$   6           &   31$\pm$   5           &                  \\
                         &            &             &                         &   15$\pm$   2           &   57$\pm$   9           &   49$\pm$   6           &   31$\pm$   5           &                  \\
                         &            &             &                         &   15$\pm$   2           &   57$\pm$   8           &   49$\pm$   6           &   31$\pm$   5           &                  \\
ESO146-G14/diffuse       &            &             &                         &   30$\pm$   8           &   93$\pm$  14           &   41$\pm$  11           &   $<$31                 &                  \\
                         &            &             &                         &   30$\pm$   8           &   93$\pm$  14           &   41$\pm$  11           &   $<$31                 &                  \\
                         &            &             &                         &   30$\pm$   8           &   91$\pm$  14           &   37$\pm$  11           &   $<$31                 &                  \\

\hline
\end{tabular}
\end{center}
{\bf  Extended Data  Table 1  $|$ PACS  and SPIRE  photometry  for the
selected  regions:}  For  each region,  we  give three  types of  flux
measurements (referred as m1, m2  and m3 in the text).  The 1-$\sigma$
flux errors  include photon  and confusion noise,  scatter of  the sky
brightness,  and uncertainties  in the  flux due  to  mis-centering of
extraction  apertures.  The  3-$\sigma$ upper  limits are  given where
appropriate.  The  uncertainties in the absolute  flux calibration are
not included here,  but are added in quadrature  before performing the
SED fitting as described in the text.
\end{table*}

\begin{table*}
\footnotesize
\begin{center}
\begin{tabular}{llllllllllllll}
\hline
region   &    f(3.6$\mu$m)    & f(4.5$\mu$m) & f(5.8$\mu$m) & f(8.0$\mu$m) & f(24$\mu$m) \\
         &    (mJy)           & (mJy)        & (mJy)        & (mJy)        & (mJy)\\ 
\hline 

SextansA/disk       &255.33$\pm$0.06     &157.29$\pm$0.05     &108.28$\pm$0.26     &59.46$\pm$0.23      &28.98$\pm$3.00      \\
SextansA/sf-1       &1.67$\pm$0.01       &1.05$\pm$0.01       &   $<$0.12          &0.62$\pm$0.04       &1.09$\pm$0.14       \\
SextansA/sf-2       &1.68$\pm$0.01       &1.36$\pm$0.01       &0.70$\pm$0.04       &0.48$\pm$0.04       &3.25$\pm$0.34       \\
SextansA/sf-3       &2.85$\pm$0.01       &2.20$\pm$0.01       &1.14$\pm$0.06       &0.60$\pm$0.05       &6.36$\pm$0.65       \\
SextansA/sf-4       &1.17$\pm$0.01       &0.69$\pm$0.01       &0.34$\pm$0.04       &0.28$\pm$0.04       &0.97$\pm$0.13       \\
SextansA/diff-1     &1.60$\pm$0.01       &1.01$\pm$0.01       &   $<$0.15          &0.81$\pm$0.04       &2.27$\pm$0.25       \\
SextansA/diff-2     &11.96$\pm$0.01      &7.43$\pm$0.01       &5.83$\pm$0.03       &2.81$\pm$0.03       &1.11$\pm$0.13       \\
SextansA/diff-3     &0.98$\pm$0.01       &0.78$\pm$0.01       &   $<$0.12          &0.93$\pm$0.04       &1.33$\pm$0.16       \\
SextansA/diffuse    &247.96$\pm$0.06     &151.99$\pm$0.05     &106.00$\pm$0.29     &57.47$\pm$0.26      &17.31$\pm$3.11      \\
ESO146-G14/disk     &4.67$\pm$0.01       &3.05$\pm$0.02       &3.00$\pm$0.08       &3.00$\pm$0.08       &3.87$\pm$0.66       \\
ESO146-G14/sf-1     &0.33$\pm$0.00       &0.24$\pm$0.00       &0.25$\pm$0.02       &0.29$\pm$0.02       &1.03$\pm$0.16       \\
ESO146-G14/sf-2     &0.38$\pm$0.00       &0.28$\pm$0.00       &0.35$\pm$0.02       &0.29$\pm$0.02       &1.23$\pm$0.18       \\
ESO146-G14/sf-3     &0.74$\pm$0.00       &0.50$\pm$0.00       &0.44$\pm$0.02       &0.49$\pm$0.02       &0.92$\pm$0.16       \\
ESO146-G14/diffuse  &3.22$\pm$0.01       &2.03$\pm$0.02       &1.96$\pm$0.08       &1.92$\pm$0.08       &   $<$2.17          \\

\hline
\end{tabular}
\end{center}

{\bf Extended Data  Table 2 $|$ Spitzer photometry:}  The {\it Spitzer}
photometric measurements were  performed in a similar way  to the {\it
Herschel} m1 method.
\end{table*}

\begin{table*}
\footnotesize
\begin{center}
\begin{tabular}{llllllllllllll}
\hline
 galaxy/band        & \multicolumn{3}{c}{Extended Source}   &   &  \multicolumn{2}{c}{Point Sources}  \\  
\cline{2-4}     \cline{6-7} \\
                    & $\sigma_{\rm measured-sky}$  &  $\sigma_{\rm HSPOT,photon}$ &  $\sigma_{\rm HSPOT,confusion}$ & &  $\sigma_{\rm HSPOT,photon}$ &  $\sigma_{\rm HSPOT,confusion}$  \\
\hline
                    &  (MJy/sr)         &   (MJy/sr)                 &     (MJy/sr)                 & &  (mJy)                 &     (mJy)             \\
\hline 
SextansA/70$\mu$m    & 2.86  & 2.03  & 0.22 &  &  0.52 & 0.08 \\
SextansA/160$\mu$m   & 1.20  & 0.92  & 0.74 &  &  0.83 & 1.34 \\
SextansA/250$\mu$m   & 0.93  & 0.24  & 1.19 &  &  2.86 & 7.0  \\
SextansA/350$\mu$m   & 0.49  & 0.11  & 0.67 &  &  2.38 & 8.2  \\
\hline
ESO146-G14/70$\mu$m  & 1.82  & 1.53   & 0.20 &  & 0.60 & 0.08   \\
ESO146-G14/160$\mu$m & 1.10  & 0.99   & 0.74 &  & 1.33 & 1.33 \\
ESO146-G14/250$\mu$m & 0.75  & 0.24   & 1.18 &  & 2.86 & 7.0  \\
ESO146-G14/350$\mu$m & 0.46  & 0.11   & 0.67 &  & 2.38 & 8.1  \\
\hline
\end{tabular}
\end{center}
{\bf  Extended  Data Table  3  $|$  Measured  sky noises of our observations 
compared  to predictions by HSPOT:} HSPOT  stands for the {\it Herschel} Observation
planning tool.
\end{table*}

\begin{table*}
\footnotesize
\begin{center}
\begin{tabular}{llllllllllllll}
\hline

region         & U$_{\rm min}$ &  U$_{\rm max}$(fixed) & $\gamma$ & $\chi^{2}$/dof & $M_{\rm dust}$ &  $M_{\rm HI}$/$M_{\rm dust}$  & $T_{\rm dust}$ \\
               &              &                      &           &               & (M$_{\odot}$)  &                           & (K)   \\
\hline

SextansA/disk        &    2.0   & 10$^{6}$ &     0.01 &     1.31 & (9.5$^{+1.1}_{-1.0}$)x10$^{3}$      & (5.7$^{+0.6}_{-0.7}$)x10$^{3}$      & 33$\pm$  1    \\
SextansA/sf-1        &    1.2   & 10$^{6}$ &     0.00 &     9.00 & (9.9$^{+2.5}_{-1.5}$)x10$^{2}$      & (1.3$^{+0.6}_{-0.7}$)x10$^{3}$      & 45$\pm$  7    \\
SextansA/sf-2        &    1.2   & 10$^{6}$ &     0.00 &    14.41 & (2.$^{+0.2}_{-0.2}$)x10$^{3}$       & (1.3$^{+0.6}_{-0.7}$)x10$^{3}$      & 28$\pm$  2    \\
SextansA/sf-3        &    4.0   & 10$^{6}$ &     0.00 &     2.87 & (1.8$^{+0.4}_{-0.3}$)x10$^{3}$      & (3.2$^{+0.6}_{-0.7}$)x10$^{3}$      & 38$\pm$  3    \\
SextansA/sf-4        &    0.7   & 10$^{6}$ &     0.00 &     2.21 & (1.6$^{+0.1}_{-0.1}$)x10$^{3}$      & (4.1$^{+5.8}_{-6.6}$)x10$^{2}$      & 27$\pm$  2    \\
SextansA/diff-1      &    4.0   & 10$^{6}$ &     0.01 &     2.07 & (5.1$^{+0.8}_{-0.5}$)x10$^{2}$      & (6.9$^{+0.6}_{-0.7}$)x10$^{3}$      & 40$\pm$  4    \\
SextansA/diff-2      &    5.0   & 10$^{6}$ &     0.00 &     0.14 & (1.8$^{+0.3}_{-0.3}$)x10$^{2}$      & (8.6$^{+0.6}_{-0.7}$)x10$^{3}$      & 30$\pm$  8    \\
SextansA/diff-3      &    4.0   & 10$^{6}$ &     0.01 &     7.20 & (3.2$^{+0.6}_{-0.3}$)x10$^{2}$      & (6.6$^{+0.6}_{-0.7}$)x10$^{3}$      & 37$\pm$  6    \\
SextansA/diffuse     &    2.5   & 10$^{6}$ &     0.01 &     0.05 & (3.1$^{+0.3}_{-0.4}$)x10$^{3}$      & (1.4$^{+0.1}_{-0.1}$)x10$^{4}$      & 29$\pm$  3    \\
ESO146-G14/disk      &    1.5   & 10$^{6}$ &     0.01 &     4.13 & (5.9$^{+0.9}_{-0.5}$)x10$^{5}$      & (2.5$^{+0.2}_{-0.5}$)x10$^{3}$      & 30$\pm$  1    \\
ESO146-G14/sf-1      &    2.5   & 10$^{6}$ &     0.01 &     3.45 & (7.5$^{+2.1}_{-1.0}$)x10$^{4}$      & (1.6$^{+0.2}_{-0.5}$)x10$^{3}$      & 44$\pm$ 12    \\
ESO146-G14/sf-2      &    4.0   & 10$^{6}$ &     0.01 &     0.19 & (6.2$^{+0.9}_{-0.8}$)x10$^{4}$      & (2.7$^{+0.2}_{-0.5}$)x10$^{3}$      & 31$\pm$  5    \\
ESO146-G14/sf-3      &    0.7   & 10$^{6}$ &     0.01 &     3.65 & (2.7$^{+0.2}_{-0.2}$)x10$^{5}$      & (5.$^{+2.2}_{-4.7}$)x10$^{2}$       & 28$\pm$  4    \\
ESO146-G14/diffuse   &    0.7   & 10$^{6}$ &     0.12 &     1.77 & (2.5$^{+0.3}_{-0.3}$)x10$^{5}$      & (4.4$^{+0.2}_{-0.5}$)x10$^{3}$      & 25$\pm$  7    \\

\hline
\end{tabular}
\end{center}
{\bf  Extended Data  Table 4  $|$  The fitting  results:} Key  derived
parameters from fitting the dust  model to the m1 flux measurements of
Extended Data Table 1 and 2.   In addition to the flux errors reported
in  Extended Data  Table 1,  the  uncertainties in  the aboslute  flux
calibration are added before performing  the fits, as described in the
text. The  last column  is the dust  temperature as given  by modified
black-body fitting.
\end{table*}

\begin{table*}
\begin{center}
\begin{tabular}{llllllllllllll}
\hline
region         &   log$\Sigma_{\rm gas}^{\rm m1-MW}$  &  log$\Sigma_{\rm gas}^{\rm m2-MW}$  &  log$\Sigma_{\rm gas}^{\rm m3-MW}$  & 
                   log$\Sigma_{\rm gas}^{\rm m1-SMC}$  & log$\Sigma_{\rm gas}^{\rm m1-LMC2}$   \\
               &   (logM$_{\odot}$/pc$^{2}$)   &    (logM$_{\odot}$/pc$^{2}$)
               &   (logM$_{\odot}$/pc$^{2}$)   &    (logM$_{\odot}$/pc$^{2}$)  &    (logM$_{\odot}$/pc$^{2}$)  \\
\hline

SextansA/sf-1    & 2.26$^{+0.23}_{-0.22}$   & 2.10$^{+0.23}_{-0.22}$   & 2.39$^{+0.24}_{-0.22}$   & 2.24$^{+0.44}_{-0.22}$   & 2.19$^{+0.57}_{-0.22}$   & \\
SextansA/sf-2    & 2.57$^{+0.21}_{-0.21}$   & 2.40$^{+0.22}_{-0.21}$   & 2.71$^{+0.22}_{-0.22}$   & 2.62$^{+0.22}_{-0.22}$   & 2.65$^{+0.21}_{-0.21}$   & \\
SextansA/sf-3    & 2.21$^{+0.23}_{-0.21}$   & 2.05$^{+0.23}_{-0.21}$   & 2.35$^{+0.22}_{-0.22}$   & 2.25$^{+0.22}_{-0.21}$   & 2.31$^{+0.22}_{-0.22}$   & \\
SextansA/sf-4    & 2.46$^{+0.21}_{-0.21}$   & 2.30$^{+0.21}_{-0.21}$   & 2.59$^{+0.21}_{-0.22}$   & 2.52$^{+0.23}_{-0.22}$   & 2.55$^{+0.22}_{-0.22}$   & \\
ESO146-G14/sf-1  & 1.21$^{+0.24}_{-0.22}$   & 1.21$^{+0.23}_{-0.22}$   & 1.25$^{+0.24}_{-0.22}$   & 1.14$^{+0.23}_{-0.22}$   & 1.15$^{+0.23}_{-0.24}$   & \\
ESO146-G14/sf-2  & 1.12$^{+0.22}_{-0.22}$   & 1.12$^{+0.23}_{-0.21}$   & 1.16$^{+0.23}_{-0.21}$   & 1.09$^{+0.22}_{-0.22}$   & 1.16$^{+0.22}_{-0.24}$   & \\
ESO146-G14/sf-3  & 1.77$^{+0.21}_{-0.21}$   & 1.77$^{+0.22}_{-0.21}$   & 1.80$^{+0.22}_{-0.21}$   & 1.82$^{+0.22}_{-0.22}$   & 1.80$^{+0.22}_{-0.24}$   & \\
\hline
\end{tabular}
\end{center}
{\bf Extended Data Table 5 $|$ Gas mass surface densities by models of
different  dust types:} Gas  surface densities  were derived  from the
dust  masses that  are  based on  IR  SED fitting  by  dust models  of
Milky-Way  (MW), Small  Magellanic  cloud (SMC)  and Large  Magellanic
cloud (LMC) grains.
\end{table*}

\begin{table*}
\begin{center}
\begin{tabular}{llllllllllllll}
\hline
region         &   $I_{\rm CO}$  &  $f_{H_{2}}$($S$(1)-17.035$\mu$m)  \\
               & (K km/s)       &  (W m$^{-2}$) \\
\hline

SextansA/sf-1      & 0.33    &  2.0E-17 \\
SextansA/sf-2      & 0.67    &  1.5E-16 \\
SextansA/sf-3      & 0.25    &  6.1E-16 \\
SextansA/sf-4      & 0.56    &  1.1E-16 \\
ESO146-G14/sf-1    & 0.02    &  4.2E-18 \\
ESO146-G14/sf-2    & 0.01    &  1.1E-17 \\
ESO146-G14/sf-3    & 0.10    &  5.4E-18 \\

\hline
\end{tabular}
\begin{center}
{\bf Extended Data Table 6 $|$ Predicted CO and warm H$_{2}$ line fluxes.}

\end{center}
\end{center}
\end{table*}

\begin{figure*}
\centerline{\includegraphics[width=1.0\textwidth]{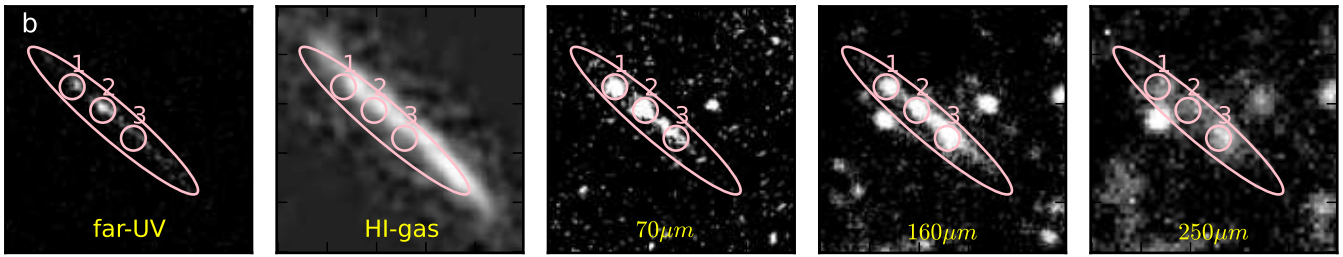} }
\vspace{-4mm}
\caption{ {\bf Mutli-wavelength  images  of  two
galaxies:} {\bf  a}, Images  of Sextans  A in the  far-UV, HI  gas, 70
$\mu$m, 160  $\mu$m and 250  $\mu$m dust emission,  respectively.  The
large circle is the  star-forming disk, small circles are star-forming
clumps  while ellipses  are diffuse  regions. {\bf  b}, Images  of ESO
146-G14 in  the far-UV, HI gas,  70 $\mu$m, 160 $\mu$m  and 250 $\mu$m
dust emission,  respectively.  The  large ellipse is  the star-forming
disk and small circles are individual star-forming clumps. }
\vspace{-4mm}
\end{figure*}






%% file: Shi_metalpoor.bbl
\begin{thebibliography}{10}
\expandafter\ifx\csname url\endcsname\relax
  \def\url#1{\texttt{#1}}\fi
\expandafter\ifx\csname urlprefix\endcsname\relax\def\urlprefix{URL }\fi
\providecommand{\bibinfo}[2]{#2}
\providecommand{\eprint}[2][]{\url{#2}}


\bibitem{Ostriker10} 
\bibinfo{author}{{Ostriker}, E. C. McKee, C. F., Leroy, A. K.}
\newblock \bibinfo{title}{{Regulation of star formation rates in multiphase galactic disks:  a thermal/dynamical equilibrium model}}.
\newblock \emph{\bibinfo{journal}{\apj}} \textbf{\bibinfo{volume}{721}},
  \bibinfo{pages}{975-994} (\bibinfo{year}{2010}).


\bibitem{Krumholz13}
\bibinfo{author}{{Krumholz}, M R.}
\newblock \bibinfo{title}{{The star formation law in molecule-poor galaxies}}.
\newblock \emph{\bibinfo{journal}{\mnras}} \textbf{\bibinfo{volume}{436}},
  \bibinfo{pages}{2747-2762} (\bibinfo{year}{2013}).

\bibitem{Bigiel08}
\bibinfo{author}{{Bigiel}} \emph{et~al.} 
\newblock \bibinfo{title}{{The star formation law in nearby galaxies on sub-kpc scales}}.  
\newblock \emph{\bibinfo{journal}{\aj}} \textbf{\bibinfo{volume}{136}},
  \bibinfo{pages}{2846-2871} (\bibinfo{year}{2008}).

\bibitem{Bolatto11}
\bibinfo{author}{{Bolatto}, A. D. } \emph{et~al.} 
\newblock \bibinfo{title}{{The state of the gas and the relation between gas and star formation 
at low metallicity: the small magellanic cloud}}. 
\newblock \emph{\bibinfo{journal}{\apj}} \textbf{\bibinfo{volume}{741}},
  \bibinfo{pages}{12-30} (\bibinfo{year}{2011}).


\bibitem{Elmegreen13}
\bibinfo{author}{{Elmegreen}, B. G.} \emph{et~al.}
\newblock \bibinfo{title}{{Carbon monoxide in clouds at low metallicity in the  dwarf irregular galaxy WLM}}.
\newblock \emph{\bibinfo{journal}{\nat}} \textbf{\bibinfo{volume}{495}},
  \bibinfo{pages}{487-489} (\bibinfo{year}{2013}).



\bibitem{Bolatto13}
\bibinfo{author}{{Bolatto}, A.} \emph{et~al.}
\newblock \bibinfo{title}{{The CO-to-H2 Conversion Factor}}.
\newblock \emph{\bibinfo{journal}{\araa}} \textbf{\bibinfo{volume}{51}},
  \bibinfo{pages}{207-268} (\bibinfo{year}{2013}).

\bibitem{Leroy11}
\bibinfo{author}{{Leroy}, A. K.} \emph{et~al.}
\newblock \bibinfo{title}{{The CO-to-H2 conversion factor from infrared dust emission across the Local Group}}.
\newblock \emph{\bibinfo{journal}{\apj}} \textbf{\bibinfo{volume}{737}},
  \bibinfo{pages}{12-24} (\bibinfo{year}{2011}).



\bibitem{Fisher14}
\bibinfo{author}{{Fisher}, D.}\emph{et~al}
\newblock \bibinfo{title}{{The rarity of dust in metal-poor galaxies}}.
\newblock \emph{\bibinfo{journal}{\nat}} \textbf{\bibinfo{volume}{505}},
  \bibinfo{pages}{186-189} (\bibinfo{year}{2014}).


\bibitem{Hunt14}
\bibinfo{author}{{Hunt}, L, K.}\emph{et~al}
\newblock \bibinfo{title}{{ALMA observations of cool dust in a low-metallicity starburst, SBS 0335-052}}.
\newblock \emph{\bibinfo{journal}{\aap}} \textbf{\bibinfo{volume}{561}},
  \bibinfo{pages}{A49} (\bibinfo{year}{2014}).


\bibitem{Remy-Ruyer14}
\bibinfo{author}{{Remy-Ruyer}, A.}\emph{et~al}
\newblock \bibinfo{title}{{Gas-to-dust mass ratios in local galaxies over a 2 dex metallicity range}}.
\newblock \emph{\bibinfo{journal}{\aap}} \textbf{\bibinfo{volume}{563}},
  \bibinfo{pages}{A31} (\bibinfo{year}{2014}).


\bibitem{Pettini04}
\bibinfo{author}{{Pettini}, M., Pagel, B.}
\newblock \bibinfo{title}{{[OIII]/[NII] as an abundance indicator at high redshift}}.
\newblock \emph{\bibinfo{journal}{\mnras}} \textbf{\bibinfo{volume}{348}},
  \bibinfo{pages}{L59-L63} (\bibinfo{year}{2004}).


\bibitem{Kniazev05}
\bibinfo{author}{{Kniazev}, A. Y.} \emph{et~al.}
\newblock \bibinfo{title}{{Spectrophotometry of Sextans A and B: Chemical Abundances of H II Regions and Planetary Nebulae}}.
\newblock \emph{\bibinfo{journal}{\aj}} \textbf{\bibinfo{volume}{130}},
  \bibinfo{pages}{1558-1573} (\bibinfo{year}{2005}).


\bibitem{Bergvall95}
\bibinfo{author}{{Bergvall}, N., Ronnback, J. } 
\newblock \bibinfo{title}{{ESO 146-G14 A retarded disc galaxy}}.
\newblock \emph{\bibinfo{journal}{\mnras}} \textbf{\bibinfo{volume}{273}},
  \bibinfo{pages}{603-614} (\bibinfo{year}{1995}).


\bibitem{Wise12}
\bibinfo{author}{{Wise}, J.} \emph{et~al} 
\newblock \bibinfo{title}{{The Birth of a Galaxy: Primordial Metal Enrichment and Stellar Populations}}.  
\newblock \emph{\bibinfo{journal}{\apj}} \textbf{\bibinfo{volume}{745}},
  \bibinfo{pages}{50-59} (\bibinfo{year}{2012}).


\bibitem{Sandstrom13}
\bibinfo{author}{{Sandstrom}, K. M.}\emph{et~al}
\newblock \bibinfo{title}{{The CO-to-H2 Conversion Factor and Dust-to-gas Ratio on Kiloparsec Scales in Nearby Galaxies}}.
\newblock \emph{\bibinfo{journal}{\apj}} \textbf{\bibinfo{volume}{777}},
  \bibinfo{pages}{5-37} (\bibinfo{year}{2013}).



\bibitem{Poglitsch10}
\bibinfo{author}{{Poglitsch}, } \emph{et~al.}
\newblock \bibinfo{title}{{The Photodetector Array Camera and Spectrometer (PACS) on the Herschel Space Observatory}}.
\newblock \emph{\bibinfo{journal}{\aap}} \textbf{\bibinfo{volume}{518}},
  \bibinfo{pages}{L2} (\bibinfo{year}{2010}).

\bibitem{Griffin10}
\bibinfo{author}{{Griffin}, M. J.} \emph{et~al.}
\newblock \bibinfo{title}{{The Herschel-SPIRE instrument and its in-flight performance}}.
\newblock \emph{\bibinfo{journal}{\aap}} \textbf{\bibinfo{volume}{518}},
  \bibinfo{pages}{L3} (\bibinfo{year}{2010}).


\bibitem{Ott12}
\bibinfo{author}{{Ott}, J.} \emph{et~al.} 
\newblock \bibinfo{title}{{VLA-ANGST: A high-resolution HI Survey of Nearby Dwarf Galaxies}}.
\newblock \emph{\bibinfo{journal}{\aj}} \textbf{\bibinfo{volume}{144}},
  \bibinfo{pages}{123-195} (\bibinfo{year}{2012}).



\bibitem{Peters13}
\bibinfo{author}{{Peters} S. P. C.} \emph{et~al.} 
\newblock \bibinfo{title}{{The Shape of Dark Matter Halos in Edge-on Galaxies: I. Overview of HI observations}}.
\newblock 
  \bibinfo{pages}{Preprint at http://arxiv.org/abs/1303.2463} (\bibinfo{year}{2013}).


\bibitem{Draine07}
\bibinfo{author}{{Draine}, B. T., Li, A. }  
\newblock \bibinfo{title}{{Infrared emission from interstellar dust. IV. 
the silicate-graphite-pah model in the post-Spitzer era}}.
\newblock \emph{\bibinfo{journal}{\apj}} \textbf{\bibinfo{volume}{657}},
  \bibinfo{pages}{810-837} (\bibinfo{year}{2007}).



\bibitem{Draine13}
\bibinfo{author}{{Draine}, B. T.}\emph{et~al}
\newblock \bibinfo{title}{{Andromeda's Dust}}.
\newblock \emph{\bibinfo{journal}{\apj}} \textbf{\bibinfo{volume}{780}},
  \bibinfo{pages}{172-189} (\bibinfo{year}{2014}).




\bibitem{Westmoquette13}
\bibinfo{author}{{Westmoquette}, M. S.}\emph{et~al}
\newblock \bibinfo{title}{{Piecing together the puzzle of NGC 5253: abundances, kinematics and WR stars⋆}}.
\newblock \emph{\bibinfo{journal}{\aap}} \textbf{\bibinfo{volume}{550}},
  \bibinfo{pages}{88-103} (\bibinfo{year}{2013}).



\bibitem{Leroy08}
\bibinfo{author}{{Leroy}, A.}\emph{et~al}  
\newblock \bibinfo{title}{{The Star Formation Efficiency in Nearby Galaxies: Measuring Where Gas Forms Stars Effectively}}.
\newblock \emph{\bibinfo{journal}{\apj}} \textbf{\bibinfo{volume}{136}},
  \bibinfo{pages}{2782-2845} (\bibinfo{year}{2008}).



\bibitem{Daddi10}
\bibinfo{author}{{Daddi}, E. } \emph{et~al.} 
\newblock \bibinfo{title}{{Different Star Formation Laws for Disks Versus Starbursts at Low and High Redshifts}}. 
\newblock \emph{\bibinfo{journal}{\apj}} \textbf{\bibinfo{volume}{714}},
  \bibinfo{pages}{118-122} (\bibinfo{year}{2010}).



\bibitem{Cormier14}
\bibinfo{author}{{Cormier}, D. } \emph{et~al.} 
\newblock \bibinfo{title}{{The molecular gas reservoir of 6 low-metallicity galaxies from the Herschel Dwarf Galaxy Survey. A ground-based follow-up survey of CO(1-0), CO(2-1), and CO(3-2)
}}. 
\newblock \emph{\bibinfo{journal}{\aap}} \textbf{\bibinfo{volume}{564}},
  \bibinfo{pages}{A121} (\bibinfo{year}{2014}).


\bibitem{Taylor98}
\bibinfo{author}{{Taylor}, C. L., Kobulnicky, H. A., Skillman, E. D.}
\newblock \bibinfo{title}{{CO Emission in Low-Luminosity, H I-rich Galaxies}}. 
\newblock \emph{\bibinfo{journal}{\aj}} \textbf{\bibinfo{volume}{116}},
  \bibinfo{pages}{2746-2756} (\bibinfo{year}{1998}).



\bibitem{Hunt10} 
\bibinfo{author}{{Hunt}, L.} \emph{et~al.}  
\newblock \bibinfo{title}{{THE    SPITZER   VIEW    OF    LOW-METALLICITY   STAR
FORMATION.   III.  FINE-STRUCTURE   LINES,   AROMATIC  FEATURES,   AND
MOLECULES}}.          
\newblock         \emph{\bibinfo{journal}{\apj}} \textbf{\bibinfo{volume}{712}},                    
  \bibinfo{pages}{164-187} (\bibinfo{year}{2010}).


\bibitem{Kuhlen12}
\bibinfo{author}{{Kuhlen}, M.} \emph{et~al.} 
\newblock \bibinfo{title}{{ Dwarf galaxy formation with H2-regulated star formation}}. 
\newblock \emph{\bibinfo{journal}{\apj}} \textbf{\bibinfo{volume}{749}},
  \bibinfo{pages}{36-57} (\bibinfo{year}{2012}).




\end{thebibliography}

\begin{thebibliography}{10}
\setcounter{enumiv}{28}
\expandafter\ifx\csname url\endcsname\relax
  \def\url#1{\texttt{#1}}\fi
\expandafter\ifx\csname urlprefix\endcsname\relax\def\urlprefix{URL }\fi
\providecommand{\bibinfo}[2]{#2}
\providecommand{\eprint}[2][]{\url{#2}}

\bibitem{Traficante11}
\bibinfo{author}{{Traficante}, A.} \emph{et~al.}
\newblock \bibinfo{title}{{Data  reduction pipeline   for  the  Hi-GAL   survey}}.
\newblock \emph{\bibinfo{journal}{Maris}} \textbf{\bibinfo{volume}{416}},
  \bibinfo{pages}{2932-2943} (\bibinfo{year}{2011}).


\bibitem{Piazzo12}
\bibinfo{author}{{Piazzo}, L.} \emph{et~al.}
\newblock \bibinfo{title}{{Artifact  removal for GD'S map  makers by  means of  post-processing}}.
\newblock \emph{\bibinfo{journal}{WEER Trans.  on Image Processing}} \textbf{\bibinfo{volume}{21}},
  \bibinfo{pages}{3687-36} (\bibinfo{year}{2012}).

\bibitem{Morrissey07}
\bibinfo{author}{{Morrissey}, P.} \emph{et~al.}
\newblock \bibinfo{title}{{The Calibration and Data Products of GALEX}}.
\newblock \emph{\bibinfo{journal}{\apjs}} \textbf{\bibinfo{volume}{173}},
  \bibinfo{pages}{682-697} (\bibinfo{year}{2007}).



\bibitem{Dale09}
\bibinfo{author}{{Dale} D. A. }  \emph{et~al.}
\newblock \bibinfo{title}{{The Spitzer Local Volume Legacy: survey description and infrared
photometry}}.
\newblock \emph{\bibinfo{journal}{\apj}} \textbf{\bibinfo{volume}{703}},
  \bibinfo{pages}{517-556} (\bibinfo{year}{2009}).


\bibitem{Engelbracht08}
\bibinfo{author}{{Engelbracht}, C. W.}\emph{ et~al}  
\newblock \bibinfo{title}{{Metallicity Effects on Dust Properties in Starbursting Galaxies}}.
\newblock \emph{\bibinfo{journal}{\apj}} \textbf{\bibinfo{volume}{678}},
  \bibinfo{pages}{804-827} (\bibinfo{year}{2008}).


\bibitem{Bertin96}
\bibinfo{author}{{Bertin}, E.} \emph{et~al.}
\newblock \bibinfo{title}{{SExtractor:  Software for source extraction}}.
\newblock \emph{\bibinfo{journal}{\aaps}} \textbf{\bibinfo{volume}{117}},
  \bibinfo{pages}{393-404} (\bibinfo{year}{1996}).

\bibitem{Sauvage11}
\bibinfo{author}{{Sauvage}, M.}
\newblock \bibinfo{title}{{Experiments in photometric measurements of extended sources}}.
\newblock {\bibinfo{journal}{http://herschel.esac.esa.int/twiki/pub/Public/PacsCalibrationWeb/ExtSrcPhotom.pdf
}}  (\bibinfo{year}{2011}).

\bibitem{Ali11}
\bibinfo{author}{{Ali}, B.}
\newblock \bibinfo{title}{{Surface brightness comparison of PACS blue array with IRAS and
Spitzer/MIPS images}}.
\newblock {\bibinfo{journal}{https://nhscsci.ipac.caltech.edu/pacs/docs/Photometer/PICC-NHSC-TN-029.pdf}}  (\bibinfo{year}{2011}).

\bibitem{Paladini12}
\bibinfo{author}{{Paladini}, R.}\emph{ et~al}
\newblock \bibinfo{title}{{Assessment analysis of the extended emission calibration
for the PACS red channel}}.
\newblock {\bibinfo{journal}{https://nhscsci.ipac.caltech.edu/pacs/docs/Photometer/PICC-NHSC-TR-034.pdf}}  (\bibinfo{year}{2012}).

\bibitem{Paladini13}
\bibinfo{author}{{Paladini}, R.}\emph{ et~al}
\newblock \bibinfo{title}{{PACS Map-making Tools: Analysis and Benchmarking}}.
\newblock {\bibinfo{journal}{http://herschel.esac.esa.int/twiki/pub/Public/PacsCalibrationWeb/pacs$\_$mapmaking$\_$report$\_$ex$\_$sum$\_$v3.pdf}}  (\bibinfo{year}{2013}).


\bibitem{Elbaz11}
\bibinfo{author}{{Elbaz}, D.} \emph{et~al.}
\newblock \bibinfo{title}{{GOODS-Herschel: an infrared main sequence for star-forming galaxies}}.
\newblock \emph{\bibinfo{journal}{\aap}} \textbf{\bibinfo{volume}{533}},
  \bibinfo{pages}{119-144} (\bibinfo{year}{2011}).


\bibitem{Egami10}
\bibinfo{author}{{Egami}, E.} \emph{et~al.}
\newblock \bibinfo{title}{{The Herschel Lensing Survey (HLS): Overview}}.
\newblock \emph{\bibinfo{journal}{\aap}} \textbf{\bibinfo{volume}{518}},
  \bibinfo{pages}{12-16} (\bibinfo{year}{2010}).

\end{thebibliography}
